\documentclass[preprint]{aastex}
\usepackage{natbib}
\usepackage{ifthen}
\newcounter{address}
\newcommand{\latin}[1]{\textit{#1}}
\newcommand{\ie}{\latin{i.e.}}
\newcommand{\eg}{\latin{e.g.}}

\newcommand{\filtermag}[1]{{\ensuremath{\left[{#1}\right]}}}
\newcommand{\ch}[1]{\filtermag{\iraclambda{#1}}}
\newcommand{\iraclambda}[1]{\ifthenelse{#1=1}{3.5}
  {\ifthenelse{#1=2}{4.5}
    {\ifthenelse{#1=3}{5.6}
      {\ifthenelse{#1=4}{7.8}
	{XXX}}}}}
\newcommand{\Halpha}{\ensuremath{\mathrm{H}\alpha}}
\newcommand{\Hbeta}{\ensuremath{\mathrm{H}\beta}}
\newcommand{\NII}{\ensuremath{\mathrm{[N\,II]}}}
\newcommand{\OIII}{\ensuremath{\mathrm{[O\,III]}}}

\begin{document}
\title{
  Anomalously low PAH emission from low-luminosity galaxies}
\author{
  David~W.~Hogg\altaffilmark{\ref{NYU},\ref{email}},
  Christy~A.~Tremonti\altaffilmark{\ref{Steward}},
  Michael~R.~Blanton\altaffilmark{\ref{NYU}},
  Douglas~P.~Finkbeiner\altaffilmark{\ref{PUO}},
  Nikhil~Padmanabhan\altaffilmark{\ref{PUP}},
  Alejandro~D.~Quintero\altaffilmark{\ref{NYU}},
  David~J.~Schlegel\altaffilmark{\ref{PUO}},
  and
  Nicholas~Wherry\altaffilmark{\ref{NYU}}
}
\setcounter{address}{1}
\altaffiltext{\theaddress}{\stepcounter{address}\label{NYU}
Center for Cosmology and Particle Physics, Department of Physics, New
York University, 4 Washington Pl, New York, NY 10003}
\altaffiltext{\theaddress}{\stepcounter{address}\label{email}
\texttt{david.hogg@nyu.edu}}
\altaffiltext{\theaddress}{\stepcounter{address}\label{Steward}
Steward Observatory, 933 N Cherry Ave, Tucson, AZ 85721}
\altaffiltext{\theaddress}{\stepcounter{address}\label{PUO}
Princeton University Observatory, Princeton, NJ 08544}
\altaffiltext{\theaddress}{\stepcounter{address}\label{PUP}
Department of Physics, Princeton University, Princeton, NJ 08544}

\begin{abstract}
The Spitzer Space Telescope First Look Survey Infrared Array Camera
(IRAC) near and mid-infrared imaging data partially overlaps the Sloan
Digital Sky Survey (SDSS), with 313 visually selected
($r<17.6~\mathrm{mag}$) SDSS Main Sample galaxies in the overlap
region.  The \iraclambda1 and $\iraclambda4~\mathrm{\mu m}$ properties
of the galaxies are investigated in the context of their visual
properties, where the IRAC \ch1\ magnitude primarily measures
starlight, and the \ch4\ magnitude primarily measures PAH emission
from the interstellar medium.  As expected, we find a strong inverse
correlation between $\ch1-\ch4$ and visual color; galaxies red in
visual colors (``red galaxies'') tend to show very little dust and
molecular emission (low ``PAH-to-star'' ratios), and galaxies blue in
visual colors (``blue galaxies,'' \ie, star-forming galaxies) tend to
show large PAH-to-star ratios.  Red galaxies with high PAH-to-star
ratios tend to be edge-on disks reddened by dust lanes.  Simple,
visually inferred attenuation corrections bring the visual colors of
these galaxies in line with those of face-on disks; \ie, PAH emission
is closely related to attenuation-corrected, optically inferred
star-formation rates.  Blue galaxies with anomalously low PAH-to-star
ratios are all low-luminosity star-forming galaxies.  There is some
weak evidence in this sample that the deficiency in PAH emission for
these low-luminosity galaxies may be related to emission-line
metallicity.
\end{abstract}


\section{Introduction}

The Spitzer Space Telescope \citep{werner04a} is opening a new window
in the Universe, with unprecedented sensitivity, angular resolution,
and spectral coverage in the mid-infrared and far-infrared.  The
Spitzer Team has made Spitzer's first science data, its ``First Look
Survey'' (FLS), immediately public.  The extragalactic component of
the FLS is in a field studied photometrically and spectroscopically in
the visual as part of the Sloan Digital Sky Survey
\citep[SDSS;][]{york00a, abazajian04a}.  This overlap permits study of
a large sample of normal, visually selected galaxies, with known
redshifts and visual spectra, in the Spitzer bands.

One of the very exciting capabilities of Spitzer is mid-infrared
imaging with the Infrared Array Camera \citep[IRAC;][]{fazio04a}.  The
longest-wavelength channel (centered at wavelength
$\lambda=\iraclambda4~\mathrm{\mu m}$) is most sensitive to large,
positive spectral features coming from PAHs associated with
interstellar dust \citep[\eg, ][]{li01pah, lu03a, smith04a}.  Since
dust is closely associated with star-formation, mid-infrared emission
is strongly related to galaxy star formation \citep[\eg, ][]{roche91a,
roussel01a, lu03a, forsterschreiber04a, pahre04a, willner04a} and
promises to reveal a lot about galaxy formation and evolution.
Unfortunately, no large, complete samples of galaxies have been
studied to date.

The interstellar medium not only traces star formation but also
obscures it.  PAH emission is an indirect and nonlinear tracer of star
formation activity, but at the same time it is expected to be an order
of magnitude less attenuated by dust than the direct measures of young
stellar populations in the visible (emission lines) and ultraviolet
(thermal continuum).  For this reason, Spitzer observations have the
potential to measure star-formation rates more fairly and to test dust
corrections, which depend not just on dust abundance but also dust
grain properties and dust geometry \citep[\eg, ][and references
therein]{calzetti01a}.

One star-forming galaxy with apparently anomalous (termed
``extraordinary'') mid-infrared properties is the $M_B=-17.2$~mag
dwarf galaxy SBS0335$-$052, which has metallicity
$12+\log_{10}(\mathrm{O}/\mathrm{H})=7.338$ \citep{izotov99a}, about
$1/20$ Solar \citep{allendeprieto01a}. This dwarf shows a deficiency of
PAH emission \citep{houck04a} despite strong star formation activity.
Whether or not this galaxy is extraordinary, it certainly demonstrates
that galaxies will show diverse properties in the Spitzer bandpasses.

Here we look at Spitzer IRAC fluxes from normal galaxies from the SDSS
spectroscopic sample, to begin an investigation of the diversity of
mid-infrared properties and their relationships with visual
properties.

\section{SDSS data}

The SDSS data used here are well described in the SDSS Data Release 2
literature \citep[][and references therein]{abazajian04a}.  The visual
photometric measurements on the data used here are discussed in detail
elsewhere \citep{blanton03d} and only briefly described here.

The SDSS $g$, $r$, and $i$ Petrosian \citep{petrosian76a} magnitudes
are corrected for Galactic extinction with the SFD maps
\citep{schlegel98a} and $K$ corrected with the \texttt{kcorrect v3\_2}
package \citep{blanton03b}, \emph{not} to redshift $z=0$ but to
roughly the median redshift of $z=0.1$.  The rest-frame bandpasses
thus made are effectively ``blueshifted'' by a factor of 1.1 and
called $^{0.1}g$, $^{0.1}r$, and $^{0.1}i$.  Calibration is to the AB
system \citep{oke83a}.  Absolute magnitudes are computed in the
$^{0.1}i$ band in the standard way \citep{hogg99cosm}, assuming a
cosmological world model with $H_0=
70\,h~\mathrm{km\,s^{-1}\,Mpc^{-1}}$, $\Omega_M=0.3$, and
$\Omega_{\Lambda}=0.7$.

The SDSS software fits various models to each galaxy image, below we
use the axis ratio b/a for the best-fit ellipsoidal exponential model
in the observed $i$ band.

In each SDSS spectrum, \Halpha, \Hbeta, \OIII\ 5007\,\AA (hereafter
just ``\OIII''), and \NII\ 6584\,\AA (hereafter just ``\NII'') line
fluxes are measured in 20~\AA\ width intervals centered on each line.
Before the flux is computed, the best-fit two-component SED model
spectrum \citep[from][]{quintero04a} is scaled to have the same flux
continuum as the data in the vicinity of the emission line and
subtracted to leave a continuum-subtracted line spectrum.  This method
fairly accurately models the \Halpha\ absorption trough in the
continuum, although in detail it underestimates the \Halpha\ fluxes
by, typically, a few percent \citep{quintero04a}.

To each galaxy's spectrum (taken through a well-centered circular
fiber with a 3~arcsec diameter), a model of star formation history and
dust attenuation is fit, providing a visual estimate of the internal
attenuation of starlight due to that galaxy's interstellar medium.  We
have designed a special-purpose code which fits a stellar population
model to the galaxy continuum and estimates the attenuation of the
starlight, under the assumption that any galaxy star formation history
can be approximated as a sum of discrete bursts.  The library of
template spectra is composed of single stellar population models
generated by population synthesis \citep{Bruzual_and_Charlot_2003}.
The models incorporate an empirical spectral library
\citep{Le_Borgne_et_al_2003} with a wavelength coverage (3200 - 9300
\AA) and spectral resolution ($\sim3$~\AA) which is well matched to
that of the SDSS spectral data.  The templates include models of ten
different ages (0.005, 0.025, 0.1, 0.2, 0.6, 0.9, 1.4, 2.5, 5, 10~Gyr)
and three metallicities (1/5 $Z_{\sun}$, $Z_{\sun}$, and 2.5
$Z_{\sun}$).  For each galaxy the templates are transformed to the
appropriate redshift and velocity dispersion and resampled to match
the data.  A non-negative least squares is performed with dust
attenuation modeled as an additional free parameter.  The attenuation
model is that the observed flux $F_{obs}$ is related to the intrinsic
flux $F_{i}$ by $F_{obs} = F_{i} e^{-\tau_{\lambda},}$ where
$\tau_{\lambda}\propto\lambda^{-0.7}$ \citep{charlot00a}.  The fit of
the models to the data is performed from 3600 to 8500~\AA\ in the
galaxy rest frame with the emission lines masked out.  In practice,
our ability to simultaneously recover age, metallicity, and
attenuation is strongly limited by the signal-to-noise of the data.
Hence we model galaxies as single metallicity populations and select
the metallicity which yields the minimum $\chi^2$.  (While this is not
particularly physical, in practice it is not a bad assumption since
the integrated light of a galaxy tends to be dominated by the light of
its most recent stellar generation.)

\section{Spitzer data}

The FLS data, like most IRAC data, were taken as a set of hundreds of
pointings with the 256$\times$256 IRAC arrays (with
$1.2\times1.2~\mathrm{arcsec^2}$ pixels).  Each of these pointings
gets passed through the Spitzer ``basic calibrated data'' (BCD)
pipeline in which it is flatfielded and calibrated and in which cosmic
rays and other bad pixels are identified and flagged.

For each galaxy in the SDSS sample, we identified all individual IRAC
BCD images in which it appears and performed, in each one, aperture
photometry through 9.2~arcsec diameter apertures.  Background (or sky
level) for each aperture flux was determined by taking a median in an
annulus of inner radius 18~arcsec and outer radius 28~arcsec.  Images
with bad pixels or cosmic rays inside the inner aperture (according to
the Spitzer-provided mask files) were excised and the photometry of
the remaining images in each band for each galaxy was averaged
together (possible because there are multiple ``dithers'' per
pointing).  The measurements were converted to AB magnitudes
\citep{oke83a}.  We applied no $K$~corrections to the Spitzer
photometry.

\section{Results}

\begin{figure}
\plotone{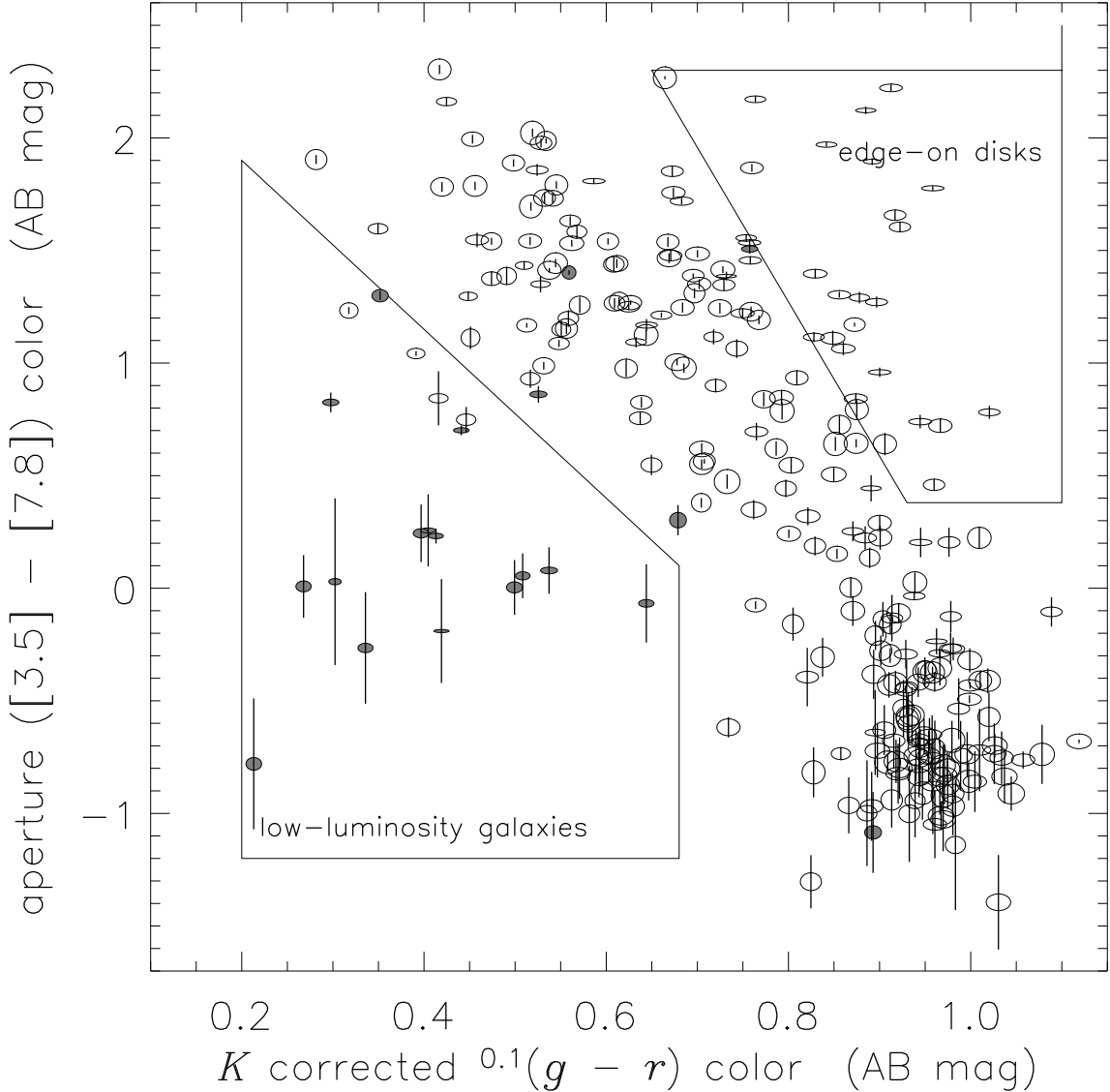}
\caption[]{Spitzer IRAC observed-frame color (aperture fluxes
explained in text) as a function of rest-frame visual color
(blueshifted SDSS bandpasses explained in text).  Symbol major axis
(horizontal extent) is linearly related to log luminosity in the
$^{0.1}i$ band (smaller symbols mean lower luminosities; compare to
Figure~\ref{fig:widget_lum}), and the symbol shape shows the galaxy
shape as observed on the sky.  Galaxies with very low luminosities
$M_{^{0.1}i}>-19$~mag are filled with grey.  Visually blue galaxies
tend to have high dust-to-star ratios and therefore red $\ch1-\ch4$
colors; visually red galaxies tend to have low dust-to-star ratios and
therefore blue $\ch1-\ch4$ colors.  Exceptions are in the outlined
regions: Red galaxies with high dust-to-star ratios tend to be edge-on
disk galaxies reddened by their own dust.  Blue galaxies with low
dust-to-star ratios tend to be low-luminosity
galaxies.\label{fig:widget}}
\end{figure}

\begin{figure}
\plotone{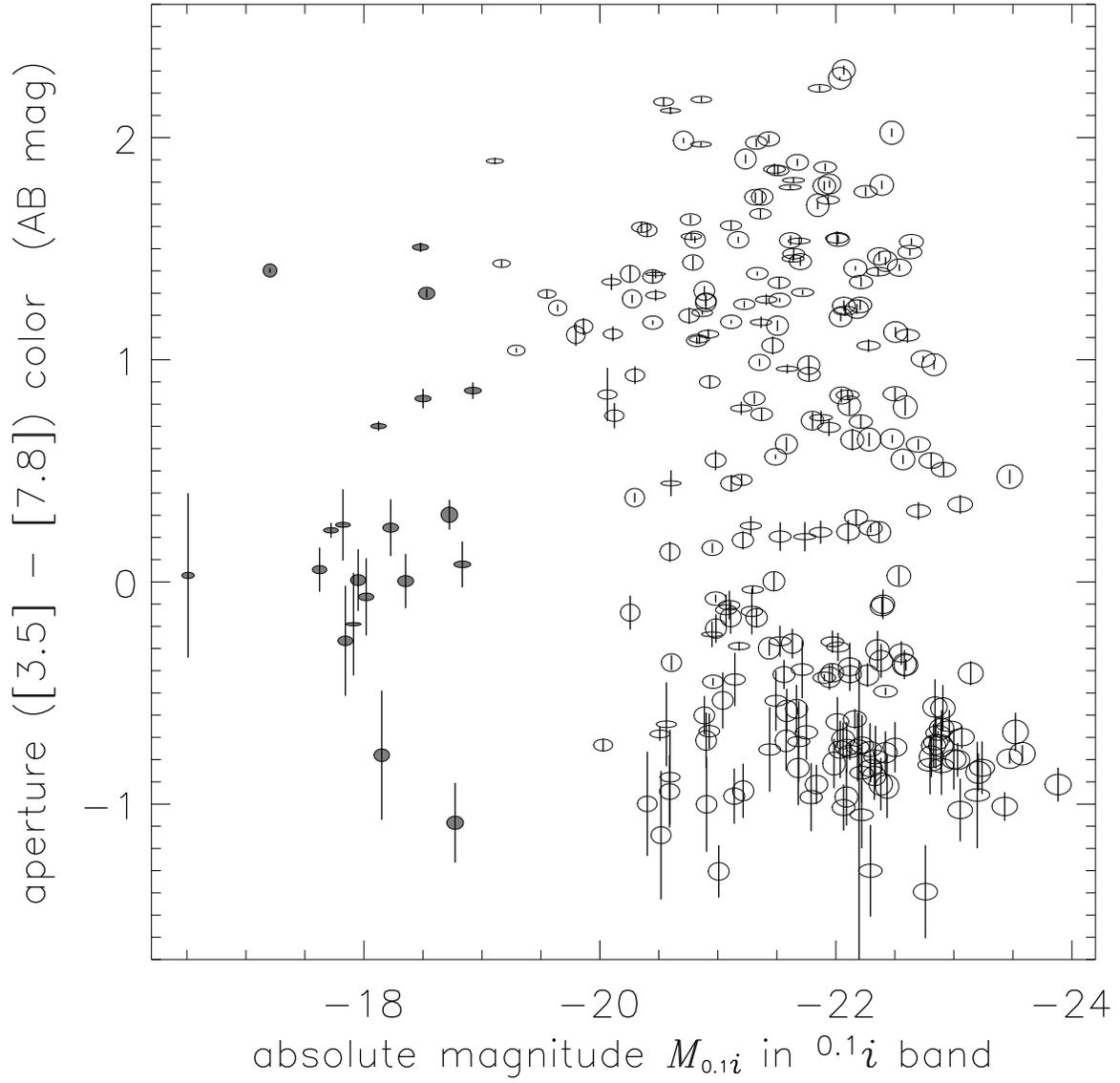}
\caption[]{Spitzer IRAC observed-frame color as a function of
rest-frame visual absolute magnitude (in the blueshifted SDSS $i$
bandpass explained in text).  Symbols as in
Figure~\ref{fig:widget}.\label{fig:widget_lum}}
\end{figure}

Figure~\ref{fig:widget} shows the observed Spitzer IRAC $\ch1-\ch4$
color as a function of rest-frame visual color, with symbol size
(major axis) linearly related to the logarithm of luminosity and shape
showing the axis ratio of the image in the $i$ band.
Figure~\ref{fig:widget_lum} shows the Spitzer color as a function of
rest-frame visual absolute magnitude (luminosity).  In
Figure~\ref{fig:widget} a clear, strong relationship between visual
color and infrared color is seen, with visually red galaxies showing
blue infrared colors and visually blue galaxies showing red infrared
colors.  This trend has a natural explanation: blue galaxies are
forming stars, star-forming galaxies contain dust, and dust produces
PAH emission features in the \ch4\ band.  Red galaxies are old and
dead; they have no material from which to make stars, therefore no
dust or PAH features.

\begin{figure}
\plotone{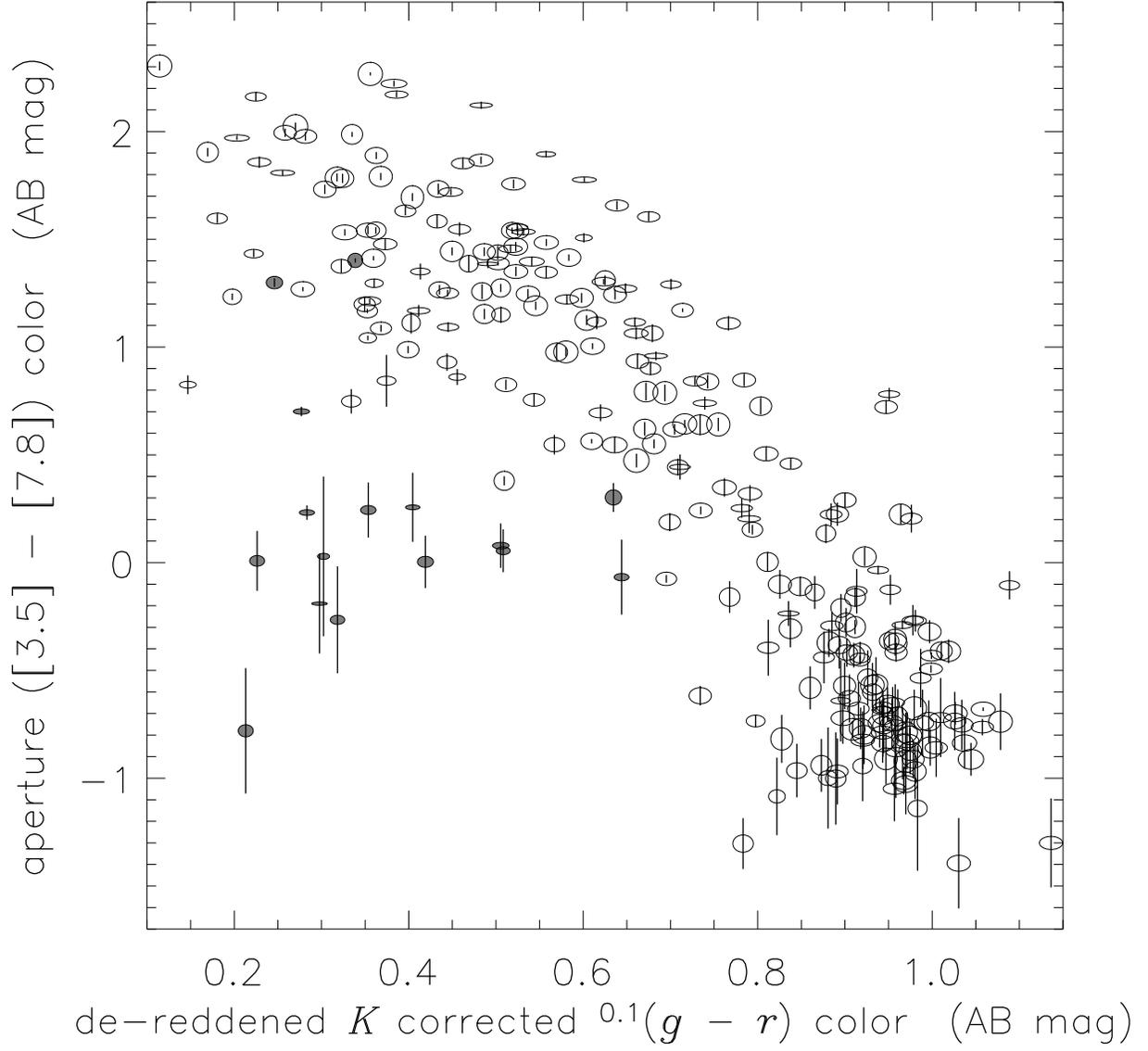}
\figcaption{Same as Figure~\ref{fig:widget} but with reddening
corrections applied (see text for details).  The edge-on galaxies are
no longer outliers from the main trend, but the low-luminosity
galaxies remain outliers.\label{fig:widget_dered}}
\end{figure}

Visually red galaxies with red infrared colors tend to be dusty,
edge-on galaxies, with high inferred visual attenuations.
Figure~\ref{fig:widget_dered} shows the same data as
Figure~\ref{fig:widget} but de-reddened assuming the best-fit
attenuation amplitude (described above) and the trivial
$\lambda^{-0.7}$ attenuation law.  This procedure is crude and not
expected to be correct in detail for any of the sample galaxies.

It is remarkable that the crude and uniform galaxy-by-galaxy
attenuation correction does a very good job of placing the visually
red but infrared blue galaxies back into the main trend in
Figure~\ref{fig:widget}.  This is a strong endorsement of the
attenuations inferred from the SDSS spectrum analysis and suggests
that simple attenuation corrections work well for normal galaxies.

The only remaining outliers on Figures~\ref{fig:widget} and
\ref{fig:widget_dered} are visually blue galaxies with
bluer-than-typical infrared colors for their visual color.  The symbol
sizes on the Figures show that these galaxies are low-luminosity
galaxies.  Evidently low-luminosity galaxies tend to be deficient in
PAH-producing (and visually extinguishing) dust.  This fits very well
with the results on SBS0335$-$052, which show a dwarf galaxy deficient
in mid-infrared PAH emission.  Indeed, it has been found in visual
observations that there is a strong relationship between a disk
galaxy's mass and its apparent interstellar medium content or state
\citep{dalcanton04a}.

\begin{figure}
\plotone{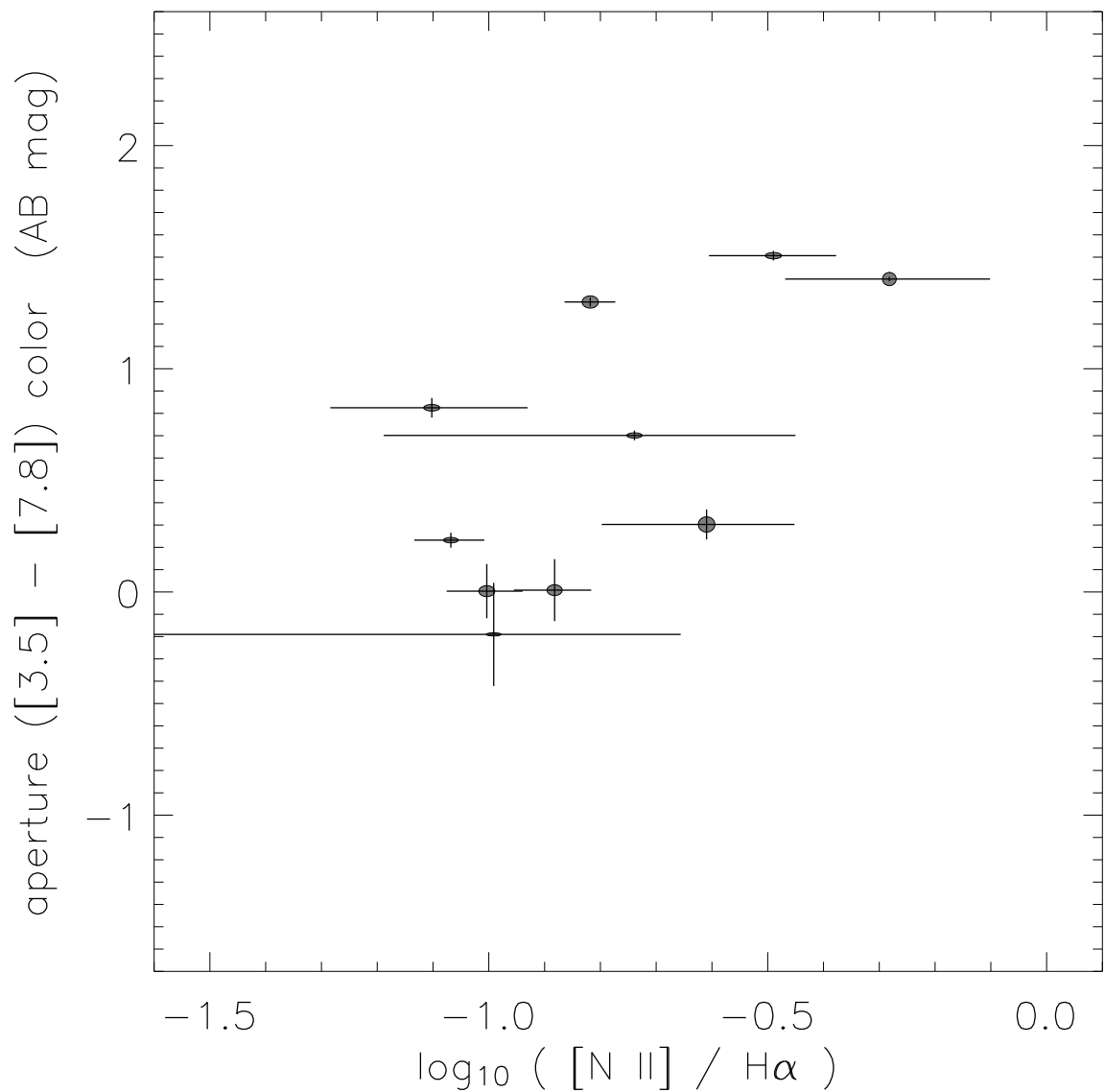} \figcaption{Spitzer
IRAC observed-frame color as a function of metallicity indicator
\NII/\Halpha\ (see text for details) for the low-luminosity
galaxies---galaxies with rest-frame visual absolute magnitude
$M_{^{0.1}i}>-19.0$---for which the ratio is measured significantly.
Symbols as in Figure~\ref{fig:widget}.\label{fig:widget_metal}}
\end{figure}

\begin{figure}
\plotone{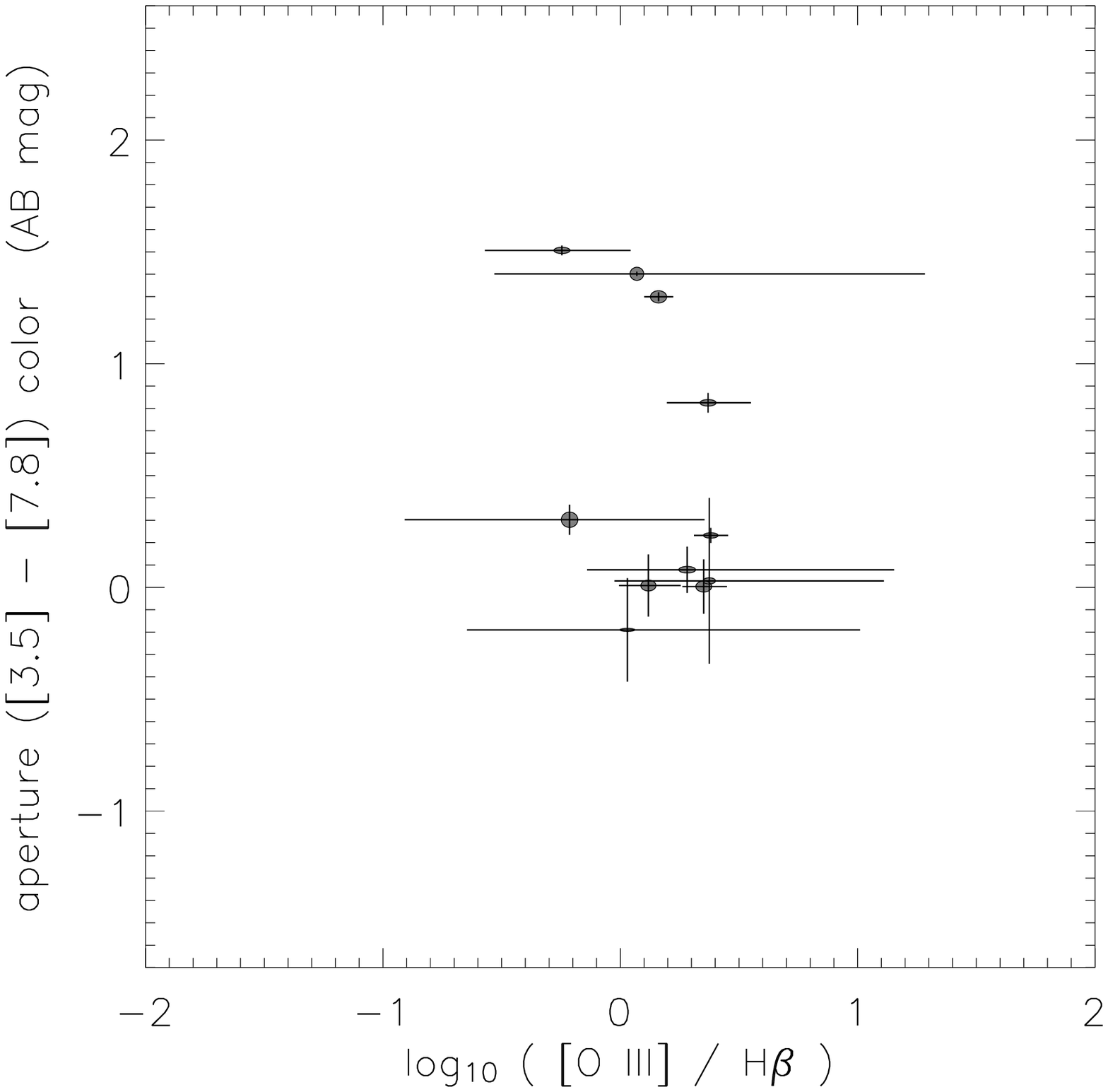} \figcaption{Spitzer
IRAC observed-frame color as a function of radiation hardness indicator
\OIII/\Hbeta\ (see text for details) for the low-luminosity
galaxies---galaxies with rest-frame visual absolute magnitude
$M_{^{0.1}i}>-19.0$---for which the ratio is measured significantly.
Symbols as in Figure~\ref{fig:widget}.\label{fig:widget_excite}}
\end{figure}

If this effect---anomalous interstellar medium content in
low-luminosity galaxies---is due to metallicity, a trend should show
up in Figure~\ref{fig:widget_metal}, which shows infrared color as a
function of line ratio \NII/\Halpha\ for the dwarf galaxies for whom
that ratio is measured to within a factor of three.  This line ratio
is a metallicity indicator with
\begin{equation}
12+\log_{10}\left(\frac{\mathrm{O}}{\mathrm{H}}\right)
 \approx 9.12 + 0.73\,\log_{10}\left(\frac{\NII}{\Halpha}\right)
\end{equation}
\citep{denicolo02a, kewley02a}.  While it is tempting to interpret the
Figure as showing a dependency of PAH emission on metallicity, given
the small sample and large errors, it shows no iron-clad trend in this
sample.  It is worth noting that none of the galaxies in this sample
are likely to have metallicities nearly as low as that of
SBS0335$-$052.  Figure~\ref{fig:widget_excite} is similar to
Figure~\ref{fig:widget_metal}, but showing radiation field hardness
indicator \OIII/\Hbeta.  No trend is visible.

\begin{figure}
\plotone{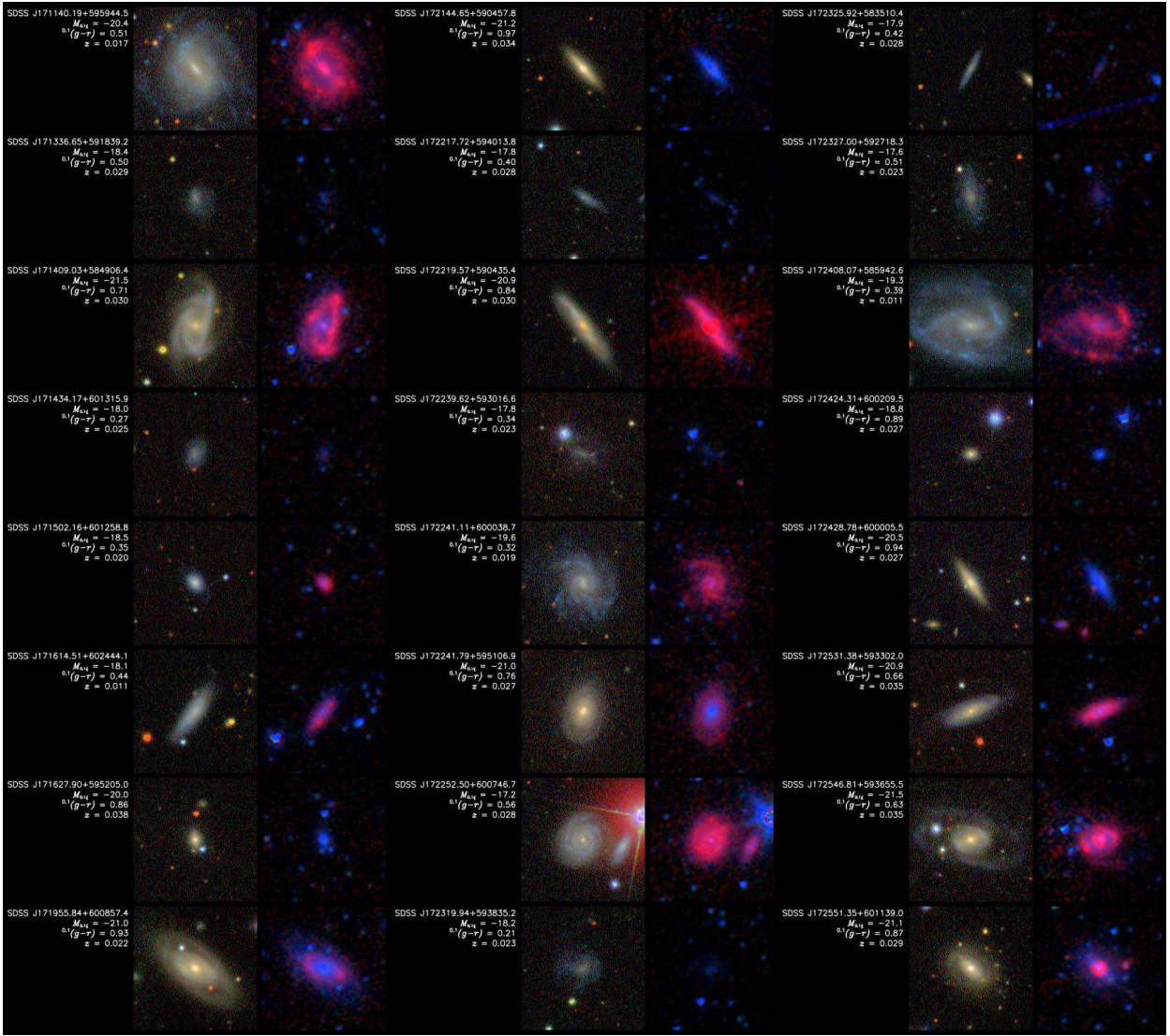}
\figcaption{Selected low-redshift galaxies from the sample.
True-color RGB images made from $73\times73~\mathrm{arcsec^2}$ cutouts
of (left) SDSS $i$, $r$, and $g$ band mosaics, and (right) Spitzer
IRAC [7.8], [4.5], and [3.5] band mosaics from the Spitzer Post-BCD
pipeline.  The images are made with identical color-preserving
stretches \citep{lupton04a}.  In the Spitzer IRAC pictures, stellar
light appears blue, and interstellar PAH emission appears
red.\label{fig:comparison}}
\end{figure}

Figure~\ref{fig:comparison} shows SDSS and IRAC images of some of the
lower redshift galaxies in the sample, which are better resolved in
the imaging, and include low luminosity members.  It is clear that for
the high luminosity galaxies, the $\ch1-\ch4$ color is a very good
predictor for attenuation-corrected visual color.  This is consistent
with the hypothesis that for these galaxies \ch4\ flux traces star
formation and \ch1\ flux traces old stellar populations
\citep{pahre04a}.  The outliers from this are all low in luminosity;
the low luminosity galaxies tend to be deficient in \ch4\ flux.

No $K$ corrections have been applied to the IRAC data.  Can the
dependence on luminosity in fact be a failure to $K$ correct (since
the low luminosity galaxies are, on average, at lower redshifts)?
There is a large, positive PAH feature in most galaxy spectra at
around $7.7~\mathrm{\mu m}$ \citep{li01pah, lu03a, smith04a} that will
tend to make higher redshift galaxies (and therefore higher luminosity
galaxies in this sample) \emph{lower} in observed-frame \ch4\ flux.
Also, the low-redshift population in this sample actually spans a
large luminosity range (see Figure~\ref{fig:comparison}), within which
the luminosity trend of PAH emission is strong.  For these reasons we
conclude that the trends we see are not due to the failure to $K$
correct.

\section{Discussion}

Luminous, star-forming galaxies are dusty, and old, red galaxies are
dust-free.  

This first look at the Spitzer First Look Survey does not show the
IRAC photometry ``diversifying'' the luminous galaxy population beyond
what is seen in the visual.  In fact, the visual color, shape,
luminosity, and spectrum of a luminous galaxy provides an extremely
good prediction for its dust content and hence its Spitzer IRAC
colors.  PAH emission is a very good tracer of star formation for
these luminous galaxies.

Low-luminosity galaxies show a deficiency in PAH emission.  They also
show much more diversity in PAH-to-star ratio; they will clearly
provide a very important subject of study for Spitzer.  Dwarfs could
be PAH deficient because they tend to be low in metallicity (\ie, lack
the material to make PAHs), because they are low in mass (\ie, lack
the gravitational potential to retain dust and molecules against
radiation pressure and winds), because they are young (\ie, have
lacked the time necessary to make PAHs), or because they have hard
internal radiation fields (\ie, have destroyed their PAHs).  Low
metallicity is the simplest explanation in many ways; indeed we have
presented weak evidence for a dependence of the PAH-to-star ratio on
metallicity among the low-luminosity galaxies.  It is also possible
that the dwarfs are not PAH deficient, but simply have different
interstellar medium geometry or different relationships between
molecules and radiation fields such that the molecules do not produce
apparent emission in the \ch4\ bandpass.

The strong, broad spectral features in the mid-IR from PAHs hold great
promise for use in photometric redshift determinations \citep[\eg,
][]{simpson99a}.  The fact that PAH emission is a strong function of
luminosity (or metallicity) may add complications to photometric
redshift schemes---because not all galaxies are drawn from the same
the same family of spectral energy distributions, and because there
might be significant families of galaxies that lack the highly
featured spectral energy distributions that are great for redshift
inference.  Of course any dependence of spectrum morphology on
luminosity can also be very useful, since inference of luminosities is
often the goal of photometric redshift determinations.  Dependencies
of spectral energy distributions on luminosity will also be important
for accurate $K$~corrections and bolometric corrections.

Because PAHs are (presumably) formed as stars evolve, the
distribution, properties, and demographics of PAHs in galaxies of
different types ought to contain information about galaxy formation
and evolution; observations of these PAHs will be part of Spitzer's
important legacy.

\acknowledgments We thank the entire Spitzer Space Telescope project
for starting the mission with a public legacy project.  We thank Aaron
Barth, Eric Bell, Julianne Dalcanton, Bruce Draine, Rob Kennicutt,
Wayne Landsman, Mike Pahre, Bill Reach, and Lisa Storrie-Lombardi for
useful data, discussions, information, or software.

The Spitzer Space Telescope is a mission of NASA.  Funding for the
creation and distribution of the SDSS has been provided by the Alfred
P. Sloan Foundation, the Participating Institutions, NASA, the NSF,
the U.S. Department of Energy, the Japanese Monbukagakusho, and the
Max Planck Society.  The Participating Institutions are The University
of Chicago, Fermilab, the IAS, the Japan Participation Group, JHU,
LANL, MPIA, MPA, NMSU, University of Pittsburgh, Princeton University,
USNO, and the University of Washington.  This research also made use
of the NASA Astrophysics Data System.  DWH, MRB, ADQ, and NW are
partially supported by NASA (grant NAG5-11669) and NSF (grant
PHY-0101738).

\bibliographystyle{apj}
\bibliography{apj-jour,ccpp}

\begin{thebibliography}{12}
\expandafter\ifx\csname natexlab\endcsname\relax\def\natexlab#1{#1}\fi

\bibitem[{Abazajian \etal(2004)}]{abazajian04a}
Abazajian, K. \etal\ 2004, \aj, in press (Data Release Two)

\bibitem[Allende Prieto, Lambert, \& Asplund(2001)]{allendeprieto01a} 
Allende Prieto, C., Lambert, D.~L., \& Asplund, M.\ 2001, \apjl, 556, L63

\bibitem[{Blanton \etal(2003{\natexlab{a}})Blanton, Brinkmann, Csabai, Doi, Eisenstein,
  Fukugita, Gunn, Hogg, \& Schlegel}]{blanton03b}
Blanton, M.~R., Brinkmann, J., Csabai, I., Doi, M., Eisenstein, D.~J.,
  Fukugita, M., Gunn, J.~E., Hogg, D.~W., \& Schlegel, D.~J. 2003{\natexlab{a}}, \aj, 125,
  2348

\bibitem[{{Blanton} \etal(2003{\natexlab{b}})}]{blanton03d}
{Blanton}, M.~R. \etal\ 2003{\natexlab{b}}, \apj, 594, 186

\bibitem[Bruzual \& Charlot(2003)]{Bruzual_and_Charlot_2003}
Bruzual, G.~\& Charlot, S., \mnras, 344, 1000

\bibitem[{Calzetti(2001)}]{calzetti01a}
Calzetti, D. 2001, \pasp, 113, 1449

\bibitem[Charlot \& Fall(2000)]{charlot00a}
Charlot, S.~\& Fall, S.~M.\ 2000, \apj, 539, 718

\bibitem[{{Dalcanton} \etal(2004){Dalcanton}, {Yoachim}, \&
  {Bernstein}}]{dalcanton04a}
{Dalcanton}, J.~J., {Yoachim}, P., \& {Bernstein}, R.~A. 2004, \apj, 608, 189

\bibitem[Denicol{\' o}, Terlevich, \& Terlevich(2002)]{denicolo02a} 
Denicol{\' o}, G., Terlevich, R., \& Terlevich, E.\ 2002, \mnras, 330, 69

\bibitem[{Fazio \etal(2004)}]{fazio04a}
Fazio, G.~G. \etal\ 2004, \apjs, in press, (astro-ph/0405616)

\bibitem[{{F{\" o}rster Schreiber} \etal(2004){F{\" o}rster Schreiber},
  {Roussel}, {Sauvage}, \& {Charmandaris}}]{forsterschreiber04a}
{F{\" o}rster Schreiber}, N.~M., {Roussel}, H., {Sauvage}, M., \&
  {Charmandaris}, V. 2004, \aap, 419, 501

\bibitem[{Hogg(1999)}]{hogg99cosm}
Hogg, D.~W. 1999, astro-ph/9905116

\bibitem[{Houck \etal(2004)}]{houck04a}
Houck, J.~R. \etal\ 2004, \apjs, in press, (astro-ph/0406150)

\bibitem[Izotov \etal(1999)]{izotov99a}
Izotov, Y.~I., Chaffee, F.~H., Foltz, C.~B., Green, R.~F., Guseva,
N.~G., \& Thuan, T.~X.\ 1999, \apj, 527, 757

\bibitem[Kewley \& Dopita(2002)]{kewley02a}
Kewley, L.~J.~\& Dopita, M.~A.\ 2002, \apjs, 142, 35 

\bibitem[Le Borgne et al.(2003)]{Le_Borgne_et_al_2003}
Le Borgne, J.~-F., Bruzual, G., Pell\'{o}, R., Lan\c{c}on, A.,
Rocca-Volmerange, B., Sanahuja, B., Schaerer, D., Soubiran, C.,
V\'{i}lchez-G\'{o}mez, R. 2003, \aap, 402, 433

\bibitem[{Li \& Draine(2001)}]{li01pah}
Li, A. \& Draine, B.~T. 2001, \apj, 554, 778

\bibitem[{{Lu} \etal(2003){Lu}, {Helou}, {Werner}, {Dinerstein}, {Dale},
  {Silbermann}, {Malhotra}, {Beichman}, \& {Jarrett}}]{lu03a}
{Lu}, N., {Helou}, G., {Werner}, M.~W., {Dinerstein}, H.~L., {Dale}, D.~A.,
  {Silbermann}, N.~A., {Malhotra}, S., {Beichman}, C.~A., \& {Jarrett}, T.~H.
  2003, \apj, 588, 199

\bibitem[{{Lupton} \etal(2004){Lupton}, {Blanton}, {Fekete}, {Hogg},
  {O'Mullane}, {Szalay}, \& {Wherry}}]{lupton04a}
{Lupton}, R., {Blanton}, M.~R., {Fekete}, G., {Hogg}, D.~W., {O'Mullane}, W.,
  {Szalay}, A., \& {Wherry}, N. 2004, \pasp, 116, 133

\bibitem[Oke \& Gunn(1983)]{oke83a}
Oke, J.~B. \& Gunn, J.~E. 1983, \apj, 266, 713 

\bibitem[{Pahre \etal(2004)}]{pahre04a}
Pahre, M.~A. \etal\ 2004, \apjs, in press, (astro-ph/0405594)

\bibitem[{{Petrosian}(1976)}]{petrosian76a}
{Petrosian}, V. 1976, \apjl, 209, L1

\bibitem[{Quintero \etal(2004)}]{quintero04a}
Quintero, A.~D. \etal\ 2004, \apj, 602, 190

\bibitem[{{Roche} \etal(1991){Roche}, {Aitken}, {Smith}, \&
  {Ward}}]{roche91a}
{Roche}, P.~F., {Aitken}, D.~K., {Smith}, C.~H., \& {Ward}, M.~J. 1991, \mnras,
  248, 606

\bibitem[{{Roussel} \etal(2001){Roussel}, {Sauvage}, {Vigroux}, \&
  {Bosma}}]{roussel01a}
{Roussel}, H., {Sauvage}, M., {Vigroux}, L., \& {Bosma}, A. 2001, \aap, 372,
  427

\bibitem[{Schlegel \etal(1998)Schlegel, Finkbeiner, \& Davis}]{schlegel98a}
Schlegel, D.~J., Finkbeiner, D.~P., \& Davis, M. 1998, \apj, 500, 525

\bibitem[Simpson \& Eisenhardt(1999)]{simpson99a}
Simpson, C. \& Eisenhardt, P. 1999, \pasp, 111, 691 

\bibitem[Smith \etal(2004)]{smith04a}
Smith, J.~D.~T. \etal\ 2004, \apjs, in press (astro-ph/0406332)

\bibitem[{Werner \etal(2004)}]{werner04a}
Werner, M.~W. \etal\ 2004, \apjs, in press (astro-ph/0406223)

\bibitem[{Willner \etal(2004)}]{willner04a}
Willner, S.~P. \etal\ 2004, \apjs, in press, (astro-ph/0405626)

\bibitem[{York \etal(2000)}]{york00a}
York, D. \etal\ 2000, \aj, 120, 1579

\end{thebibliography}

\end{document}